# Omega-blocks with spatially compounding extremes over Europe are highly sensitive to remote atmospheric drivers

## Short Title: Omega-blocks highly sensitive to remote drivers

Magdalena Mittermeier[1*], Christian M. Grams[2], Urs Beyerle[3], Laura Suarez-Gutierrez[3,4], Emanuele Bevacqua[5], Yixuan Guo[6], Jakob Zscheischler[5,7], & Erich M. Fischer[3]

[1] Department of Geography, Ludwig-Maximilians-Universität München, Munich, Germany

[2] Federal Office of Meteorology and Climatology MeteoSwiss, Zurich-Airport, Switzerland

[3] Institute for Atmospheric and Climate Science, ETH Zurich, Zurich, Switzerland

[4] Meteorology and Air Quality Group, Wageningen University & Research, Wageningen, the Netherlands
[5] Department of Compound Environmental Risks, Helmholtz Centre for Environmental Research – UFZ, Leipzig, Germany
[6] College of Oceanic and Atmospheric Sciences, Ocean University of China, Qingdao, China

[7] Department of Hydro Sciences, TUD Dresden University of Technology, Dresden, Germany

*corresponding author: m.mittermeier@lmu.de

**Abstract**

Omega-blocks can trigger spatially compounding heat-precipitation extremes with severe societal impacts, as seen in September 2023 when a heatwave over France coincided with devastating floods in the Iberian Peninsula and Greece. Although blocking in general has been linked to moist processes in upstream warm conveyor belts (WCBs), it has remained unexplored whether and how upstream WCB activity influences the evolution of omega-blocks and downstream flood-heat-flood impacts. Here, we show that already five days ahead, small differences in the upstream evolution – particularly in WCB outflow regions – distinguish cases that later produce extreme compound events over Europe from weaker ones, even though their large-scale anomalies initially appear similar. We illustrate the distinct evolution in remote locations by analyzing storylines simulated in a fully coupled climate model. Using ensemble boosting, we generate hundreds of physically plausible simulations of omega-prone situations. Lagrangian air parcel tracking reveals that variations in WCB outflow areas can explain differences in upstream precursors and downstream effects over Europe. Our results highlight ensemble boosting as a powerful approach to systematically track dynamical differences along model-based event storylines, important for understanding and anticipating compound extremes striking multiple regions simultaneously.

**Teaser**

Heat-Precipitation Extremes over Europe are highly sensitive to cyclones off the East Coast of Canada five days earlier.

**Introduction**

In September 2023, Europe experienced a set of nearly co-occurring extreme events across the continent. While Mediterranean areas were hit by extreme precipitation and floods, Central Europe experienced a heatwave at the same time. Severe flooding occurred in Spain (September 3rd) and on the other hand in Greece, Turkey, and Bulgaria (September 4-7th) with a total of 33 fatalities. Later, the second low-pressure system called "Storm Daniel" developed into a deadly/devastating Mediterranean subtropical storm (Medicane) and led to more than 11,000 confirmed casualties following dam collapses on September 10th in Libya[1-4].

These spatially compounding extremes[5-6] of opposite types (wet vs. hot) were connected by a persistent atmospheric omega-block over Europe[7], that is, a high-pressure system flanked by two low-pressure systems, forming the shape of the Greek letter Ω (see Fig. 1a). Atmospheric blocking is a major source of heatwaves over Europe[8-11]. In combination with low-pressure systems, blocking is preconditioning compound events with heavy precipitation and flooding in adjacent regions[11]. For the 2021 North American heat wave, several studies stress that the block causing the heat was affected by dynamical processes on the synoptic scale, in particular cyclones and atmospheric rivers[12] and even Rossby waves propagating from far upstream[13].

Numerous conceptual frameworks have been proposed for blocking formation and maintenance (see review by ref. 9). Blocking events typically develop when large-scale poleward advection of air masses with low potential vorticity (PV) occurs, driven by synoptic-scale eddies and a meridionally amplified flow pattern, which builds up an initial large-scale upper-level ridge[14-16]. Next, two primary pathways exist that transport air masses with low PV into the block[17]: (a) quasi-adiabatically within near-tropopause airstreams along an amplified upper-level jet, reflecting dry dynamical processes[18-20], or (b) via ascending warm conveyor belt (WCB) airstreams associated with strong latent heat release, representing moist processes[21]. While early studies[18-20,16] have mainly focused on dry dynamical processes, recent studies[21-27] have shed light on the importance of moist dynamical processes. In the WCBs of upstream cyclones, moist air masses ascend, which leads to cloud formation, condensation, and latent heat release. This generates anticyclonic PV anomalies in the area of WCB outflow, which can intensify an initially existing upper-level ridge[28] and thus contribute to blocking formation[21-27]. In addition, WCB activity can be involved in the initiation and amplification of a Rossby wave packet and downstream development, which might ultimately result in blocking far downstream of the actual WCB activity[29,30,13]. Although the relative role of dry and moist dynamics remains disputed, recent studies emphasize the high case-to-case variability as well as the high relevance of the perspective adopted by the different diagnostic frameworks[31,24]. WCB outflows further often co-occur with PV streamers and Rossby wave breaking[32], which might contribute to cut-off low formation in omega-blocking. Vice versa, it is well established that PV streamers are precursors to heavy precipitation events in Europe[33], and these might occur as a downstream impact of recurring North Atlantic tropical cyclones[34]. In general, based on a climatological analysis, the odds of heavy precipitation are increased in areas southwest to southeast of blocking systems, and in some cases also to their north[35].

While moist processes in blocking formation have been receiving more attention lately, there is still a knowledge gap on the role of moist processes for omega-blocking with related compounding heat and precipitation extremes through cut-off low formation adjacent to the block. In particular, it remains unclear whether moist processes can explain why a dynamical "event-prone"[36] situation develops omega-blocking with compounding extremes or not. This gap is highly relevant, as omega-blocking and associated compounding extremes pose severe risks. Not only due to the potentially severe individual events (e.g., flooding), but especially because of the simultaneous occurrence of multiple extreme events across the continent, which can strain management facilities such as the joint European Civil Protection Pool in the EU[37,38].

This study investigates the large-scale atmospheric conditions that determine whether an omega-block with strong, spatially compounding temperature and precipitation extremes develops. We address the knowledge gap on the role of moist processes in upstream cyclones related to omega-blocking by explicitly investigating omega-blocks within a fully coupled climate model. Our central research question is: what aspects of the large-scale circulation determine whether an omega-block with compounding extremes forms? We hypothesize that upstream cyclone activity and associated warm conveyor belts play a central role.

**Results**

**Ensemble boosting can simulate spatially compounding events similar to the observed event in September 2023**

Using ensemble boosting[39-41] (see Methods), CESM2[42] can efficiently simulate events that are remarkably similar to the observed event in September 2023 (Fig. 1). The center of the blocking anticyclone of the model-based event in Fig. 1b is slightly further to the North-West compared to ERA5 (Fig. 1a), and it shows a distinct omega-block with spatially compounding extremes over the 2023 impact boxes. This model-based event ranks highest among compound precipitation–temperature–precipitation extremes in the boosted ensemble, as identified by a novel multivariate ranking metric[43] (see Methods) employing three predefined regional boxes (Fig. 1).

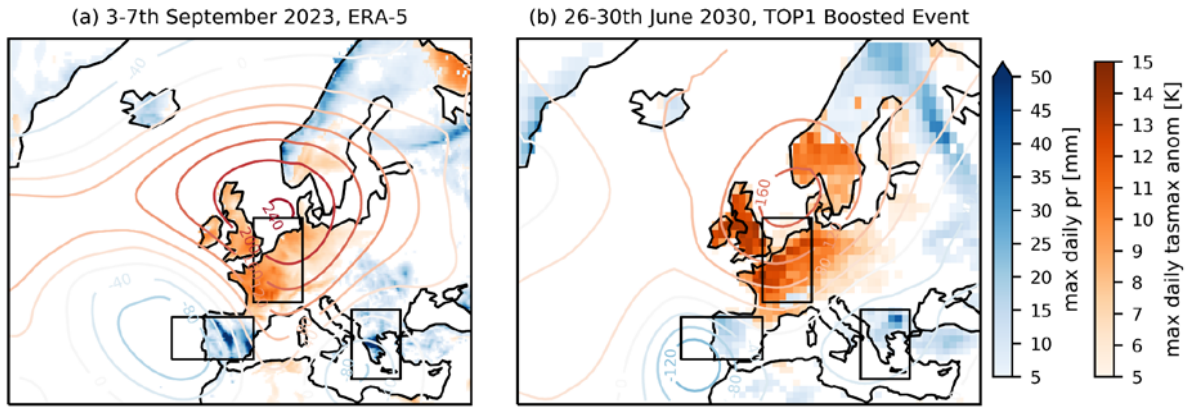

**Fig. 1 | Omega-Pattern and spatially compounding extremes during the observed and the TOP1 boosted CESM2 event.**
**a** Averaged 500 hPa geopotential height anomalies (contours), maximum of daily maximum temperature (tasmax) anomalies (red; only plotted for anomalies > 5K; threshold chosen for readability purpose), and maximum daily precipitation (only plotted for pixels with pr > 5 mm; threshold chosen for readability purpose) from 3-7th September 2023, the period of observed flooding in Spain and Greece, in ERA5. Anomalies are calculated compared to the climatological mean of 2005-2024 (31-day running mean for z500; 15-day running mean for tasmax). Temperature and precipitation anomalies are clipped with the land-sea mask. Precipitation is plotted above the temperature layer. **b** Same as in (a) but for the highest ranking event (TOP1) from the boosted ensemble of the CESM2 and the period from 26th-30th June 2030. Anomalies are calculated as for (a), but compared to the 30-member CESM2-LE 2005-2024 climatology.

The boosted ensemble reveals a wide range of possible event evolutions, demonstrating strong sensitivity to small perturbations in initial conditions. At shorter lead times, the trajectories remain dynamically close to the unperturbed reference event, producing moderate variability, whereas longer lead times allow greater divergence and more extreme outcomes[44]. However, extended lead times also increase the likelihood of losing the omega-blocking structure entirely. To retain the characteristic omega-pattern while capturing sufficient variability in compounding extremes, relatively short lead times of 10 to 18 days are used here.

Ensemble boosting[39-41] is an event-based reinitialization technique applied to an unperturbed reference event selected from the CESM2-LE (see Methods and 43 for details). The method generates hundreds of physically consistent event trajectories by introducing small perturbations to the atmospheric state several days before the event peak. Compared to approaches that boost without conditioning on specific circulation patterns[39-41,45], our setup employs shorter lead times, consistent with studies that target dynamically defined events[46].

**Top and minor compound events differ in the pre-existing upper-level ridge over Scandinavia and the Atlantic area**

Ensemble boosting yields very different event trajectories that allow us to study why some events are much more extreme than others. We find that the hottest events over France are not necessarily the wettest over the Iberian Peninsula and/or Greece, and others are very wet over the Iberian Peninsula or Greece, but not so extreme in the other two regions of interest. To systematically investigate differences in the atmospheric conditions between the 800 boosted simulations, we select the members with the 10% strongest and 10% weakest spatially compounding events in the selected regions (red/blue dots in Fig. 2). For this selection, compound events are ranked based on a metric[43] that accounts for precipitation over the Iberian Peninsula and Greece and maximum daily temperature over Northern France— that is, in difference to ref. 43, the atmospheric flow is not considered here.

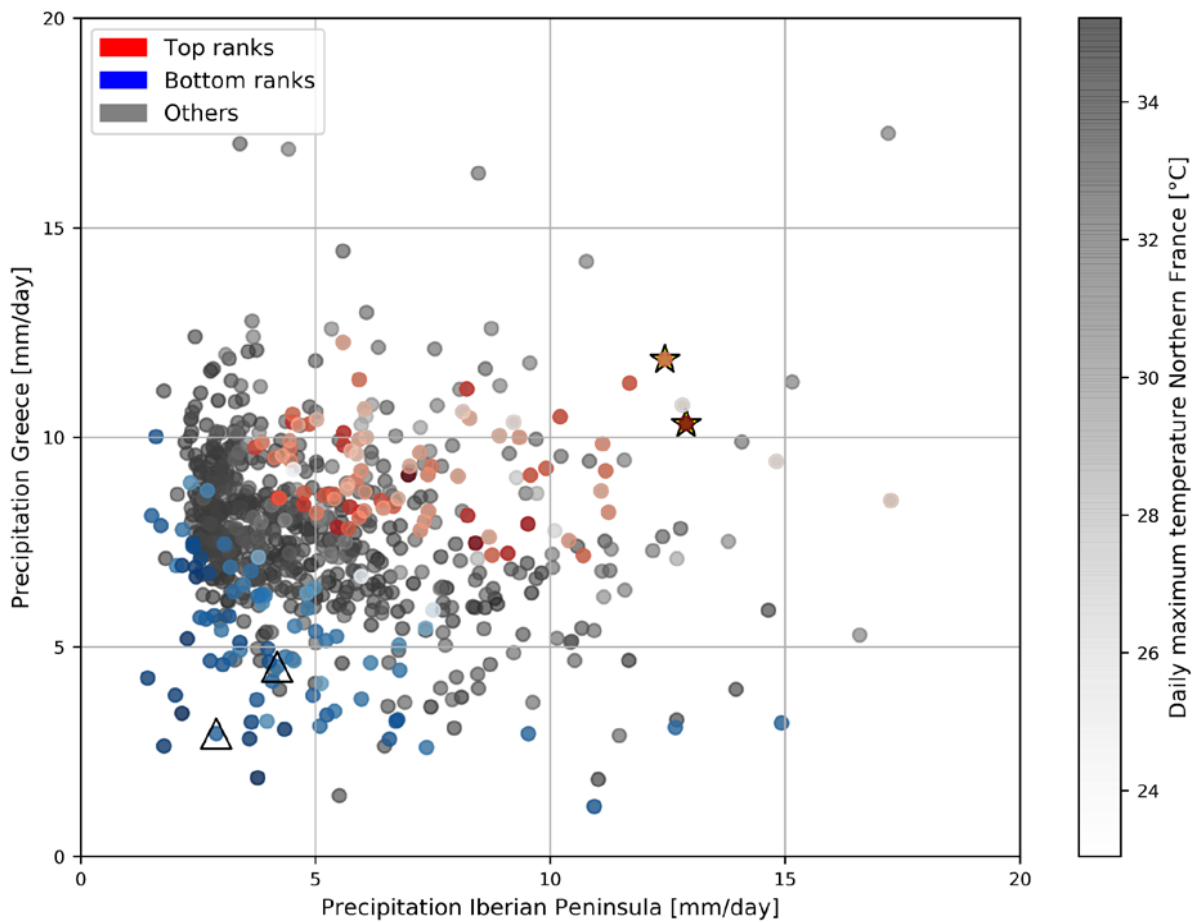

**Fig. 2 | Groups of top and minor compound events.** 5-day maximum daily precipitation [mm/day] after smoothing with 3-day running mean (5-day max. of Rx3day) averaged over the 40% wettest pixels of the Iberian Peninsula box (see Fig. 1; x-axis) and the Greece box (y-axis) and 5-day maximum of maximum daily temperature [°C] after smoothing with 3-day running mean (5-day max. of Tx3day) averaged over the 40% hottest pixels of the Northern France box (color shading; see colorbar). Red/blue dots represent the 10% of members (n=80) with the strongest/weakest compounding extremes (top/minor ranks), considering a joint ranking metric for compounding extremes (see ref. 43). The precipitation and maximum temperature values are derived from the day with the strongest compounding extremes for each member (best day with the lowest joint ranking number). The stars/triangles mark the two members each selected from the top/minor compound event group for detailed analysis below.

The top compound event group shows a distinct omega-blocking pattern on the peak date of the reference event (Fig. 3k) over Europe. This reveals that a pronounced omega-blocking plays a crucial role in driving extremes of this type of spatially compounding events in Europe. The core of the high-pressure system over Great Britain and southern Scandinavia is flanked by low-pressure systems near the eastern coast of the Iberian Peninsula and over Greece. A distinct sequence of processes is apparent already during the onset of the omega-blocking (Fig. 3a,c,e,g,i), showing a previously existing upper-level ridge over Great Britain/Southern Scandinavia. Two days before the peak date of the reference event (2030-06-26; Fig. 3g), a second eastward-moving high-pressure system merges with the pre-existing upper-level ridge. As a result, the upper-level trough between them becomes separated from the westerly flow, leading to the development of a cut-off low over the Iberian Peninsula, as indicated by the closed contours of the absolute Z500 field on 2030-06-27 (Fig. 3i).

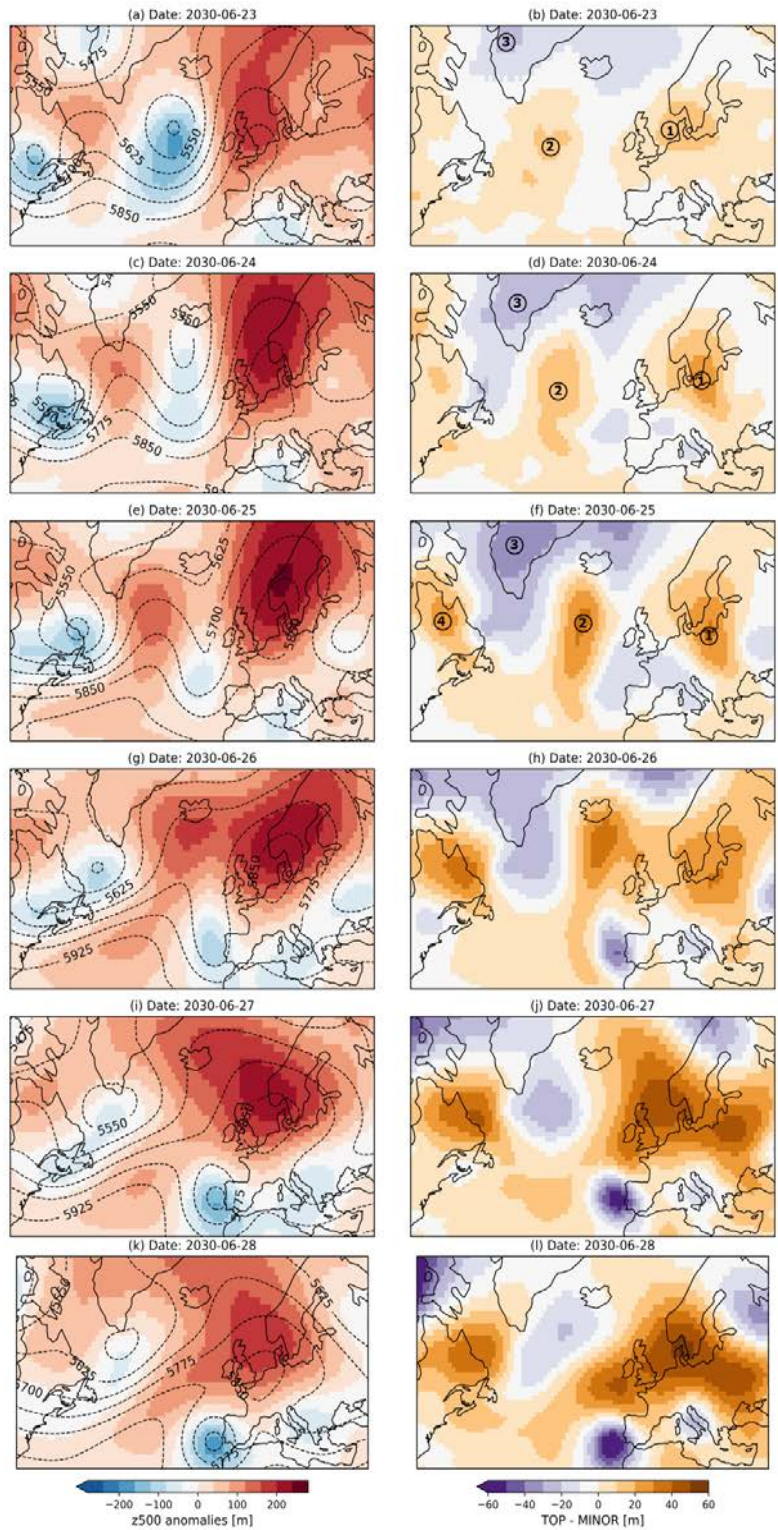

**Fig. 3 | How does the large-scale 500hPa geopotential height field differ between the top and minor compound event groups? a, c, e, g, i, k** Composites of 500hPa geopotential height values (dashed contours) and anomalies (colors) averaged over 80 members of the top group (see Fig. 2) during the onset and peak of the omega-pattern over Europe (period of 23-06-2030 to 28-06-2030). Anomalies are calculated compared to the 2005-2024 climatology of the 30-member CESM2-LE (31-day running mean) for the summer half-year (JJASON) in the mid-latitudes (50-80°N). **b, d, f, h, j, l** Difference between composites of 500hPa geopotential height anomalies of top and minor compound event groups. Key features mentioned in the text are indicated with numbers.

Five days before the peak date of the reference event (2030-06-23), the composite maps of the two groups (top and minor compound events) are very similar, with a relatively consistent large-scale dynamic pattern and pixelwise differences less than 20 m geopotential height (Fig. 3b and Fig. S1a). This similarity is in line with the minimal perturbations of initial conditions only 5-13 days before June 23rd, when the boosted simulations were started (see Fig. S2 and S3 for temporal evolution of standard deviations among boosted members). Differences are hard to see, even though the events are shaped days ahead of the peak date of the reference event (see below). Most importantly, distinct differences between the top and minor compound events emerge quickly during the following few days. Three days before the peak date of the reference event (2030-06-25, Fig. 3f), four "centers of action" with clear distinctions between the two groups are visible (highlighted with numbers in the figure): 1) positive values over southern Scandinavia/Baltic states in the vicinity of the pre-existing upper-level ridge, which is more pronounced in the top group, 2) positive values over the Atlantic south of Iceland in the transition of the upper-level ridge to the upper-level trough, 3) negative values over Greenland extending to the Labrador Sea, where z500 anomalies are slightly lower in the top group, and 4) positive values over Québec indicating higher z500 anomalies in the minor compound event group. The major differences between the groups are visible on the peak date of the reference event (2030-06-28, Fig. 3l), with a weaker omega-pattern in the minor compound event group with pixelwise differences of up to -80 meters.

Some members capture one of the extremes but not the entire impactful compound event. Thus, the spread within the minor group is higher than in the top compound event group (see Fig. S2 and Fig. S3). The majority of members of the minor group show a distinct omega pattern over Central Europe, but the spatially compounding extremes are weak in the target regions (e.g., Fig. S9). Other members show a blocking anticyclone at a different location, mostly over Iceland or Scandinavia (e.g., Fig. S8), and a marked shift of the regions affected by high temperatures or increased precipitation. Except for a few members, these displaced blocking cases come without an omega-pattern or without three spatially compounding extremes. In some members, a blocking anticyclone is accompanied by an upstream cut-off low (dipole block). In rare cases, the blocking is lost entirely (not shown). In summary, the boosting reveals a high variability in the exact characteristics of blocking and associated extremes in this event-prone large-scale setting.

**Strongest compound events are shaped days ahead over the North Atlantic**

The differences between the top and minor compound event groups in Fig. 3 suggest that variations in the atmospheric conditions in the days before the peak of the event are primarily associated with two features of the large-scale atmospheric circulation: i) the presence of an upper-level ridge over Scandinavia (see Fig. 3a,c,e), which coincides with the core area of the blocking anticyclone, and ii) a transition zone over the Atlantic, with both the upper-level ridge that later merges into the pre-existing block and the trough that later becomes the cut-off low over Iberia (see Fig. 3g). To further investigate these relationships, we calculate Pearson correlation coefficients between the large-scale atmospheric flow represented by 500hPa geopotential height anomalies and the weather on the ground, specifically temperature over Northern France, and precipitation over the Iberian Peninsula and Greece.

Boosted members with stronger high-pressure systems over the heatwave area tend to have higher temperature extremes there (Fig. 4). Specifically, we find high correlations (across ensemble members) of up to 0.82 between z500 anomalies over Northern France and maximum daily temperature averaged over the 40% hottest pixels of the Northern France box on the day of event peak (June 28th, Fig. 4f). Conversely, there is a negative correlation between the low-pressure areas of the omega-pattern and the temperature extremes over Northern France of up to -0.67. In other words, boosted members with more pronounced omega-patterns tend to also have higher temperature extremes.

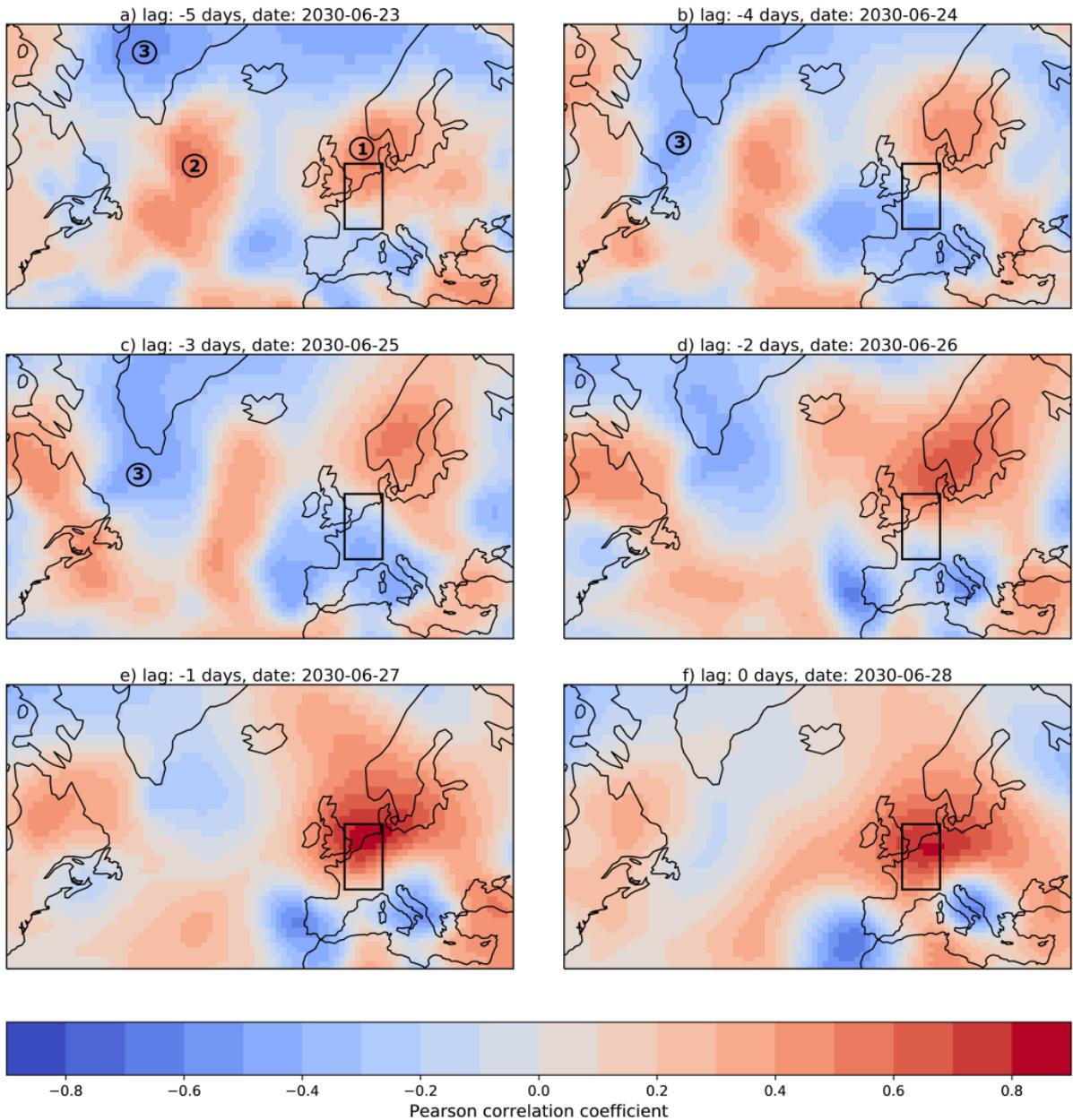

**Fig. 4 | Do we see correlations with remote areas? a, b, c, d, e, f** Pearson correlation maps between maximum daily temperature averaged over the 40% hottest pixels in Northern France (outlined box) on the event peak day of the reference event (2030-06-28), and 500 hPa geopotential height (z500) anomalies at each grid point across the domain, for various time lags: panels (a-f) z500 anomalies from 5 days before event peak to the day of event peak in the reference event (2030-06-28). Correlations are computed using a pooled dataset comprising the top 10% and bottom 10% of boosted ensemble members (n = 160). z500 anomalies are calculated relative to the 31-day running mean climatology of the 30-member CESM2-LE in the years 2005 to 2024 for the summer half-year (JJASON) in the mid-latitudes (50-80°N). Number labels mark features mentioned in the text.

High correlations are also already visible days before the event peak in three "centers of action", two of them in remote areas over the Atlantic (June 23-25th 2030, Fig. 4a-c), pointing to potential precursors. Five days before the event peak (June 23rd 2030; Fig. 4a), positive correlations >0.5 are visible: 1) over Scandinavia, the position of the pre-existing upper-level ridge, and 2) over an area over the Atlantic in the South-East of Greenland. These centers of positive correlations go in line with the differences between top and minor compound events in Fig. 3b. The second center of positive correlations (label 2) moves eastward in the following days along the westerly flow. On June 26th (Fig. 4d) positive correlations are visible in the area, where the upper-level ridge over the Atlantic merges with the one pre-existing over Scandinavia (area between Iceland and Great Britain/Scandinavia in Fig. 4d). 3) distinct negative correlations of <-0.5 are visible five days before the event peak over Greenland (see label 3 in Fig. 4a) and <-0.4 four to three days before the event peak (Fig. 4b,c) over a band from Greenland over the Labrador Sea to Newfoundland. A similar pattern emerges when looking into correlations of the z500 field with precipitation extremes over the Iberian Peninsula (see Fig. S4) and Greece (Fig. S5), even if the correlations tend to be lower here. The maps show positive correlations in an area over the Atlantic in the South-East of Greenland up to five days before the event peak, as well as positive correlations over Great Britain/Scandinavia, in the position of the pre-existing upper-level ridge. However, unlike for temperature, the correlations for precipitation are consistently below 0.5.

In summary, Fig. 4 reveals remarkably high correlations (exceeding ±0.5) between z500 anomaly fields over the Atlantic five days before the event peak and daily maximum temperature over Northern France on the event peak. Together with the differences in composite maps of top and minor compound events in Fig. 3, this is one more line of evidence for the influence of remote areas on the development of an omega-block with strong spatially compounding extremes.

**Sensitivity to far upstream precursors**

Lagrangian air parcel tracking[47] reveals a direct connection between far-upstream precursors and downstream compound extremes in Europe. In all four selected ensemble members representing the top and minor compound event groups (see Methods and markers in Fig. 2), a low-pressure system with potential for warm conveyor belt (WCB) activity was present on the east coast of Canada (Québec/Newfoundland) from June 23–25 (Fig. 3a,c,e). Forward trajectories from the low-pressure area over 48 hours show uplifting air parcels that meet the WCB criterion (uplift of ≥ 400 hPa/48 h starting from 1000–700 hPa). This is true for all four members starting on June 22nd, six days before the event peak, with a comparable number of trajectories in both the top and minor compound events. The statistical correlations between North Atlantic flow and downstream extremes (Fig. 4) thus reflect a physically coherent link to upstream synoptic conditions.

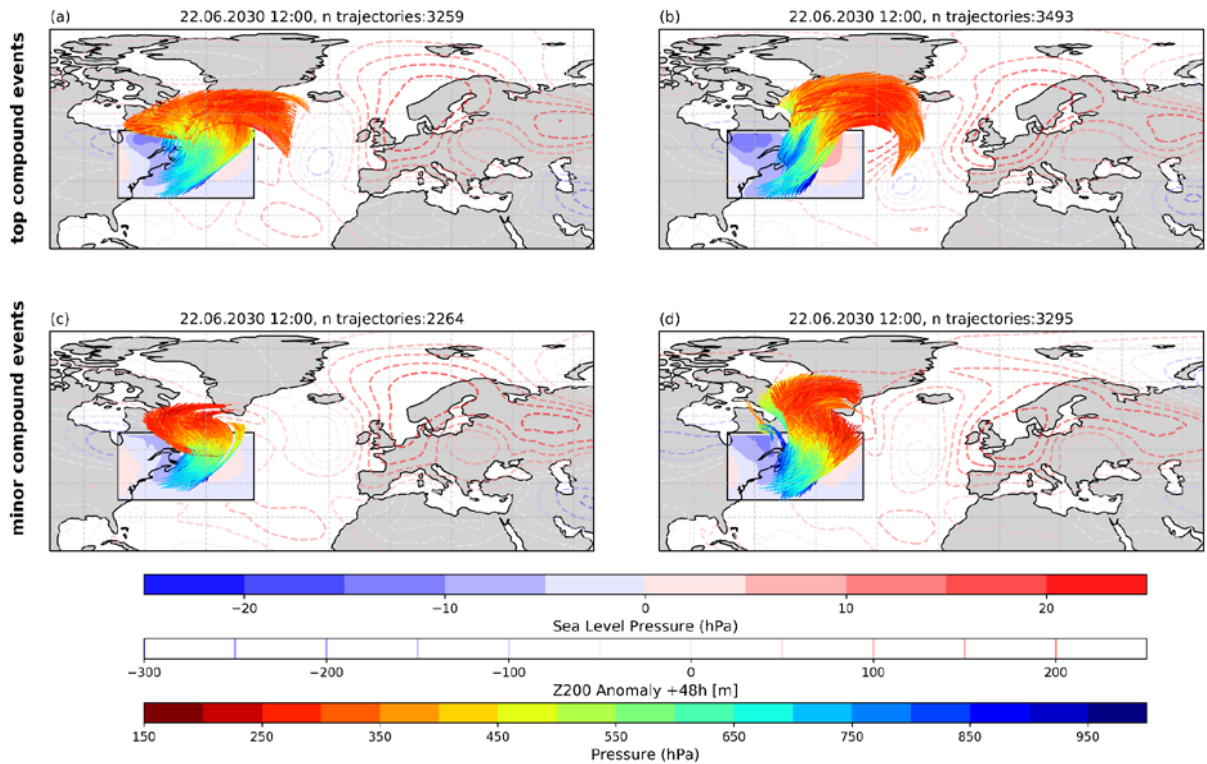

**Fig. 5 | Is warm conveyor belt activity playing a role? a, b** Pressure levels of ascending air streams within warm conveyor belts (colored trajectories) for two top compound event examples. **c, d** Same as a, b, but for two minor compound event examples. 48-h forward trajectories are started on June 22nd, 2030, at 12 am from 100 km equidistant positions within the outlined box from levels between 100-700hPa. The sea level pressure anomalies in hPa are illustrated by filled contours within the starting box. Dashed contours show 200hPa geopotential height anomalies 48h later, representing the large-scale anomaly field at the outflow areas of the warm conveyor belts.

The direction of outflow, however, is substantially different (Fig. 5). While minor compound events have a northward outflow, top events show anticyclonic wavebreaking into Europe with a warm conveyor belt outflow directed to the North-East towards the upper-level ridge located in the South of Greenland on June 24th. The trajectory frequencies per pixel averaged throughout June 20-23rd in Fig. 6 illustrate this even more. Interestingly, the density of trajectories passing a certain pixel (at levels < 450hPa) is higher for the minor compound examples (Fig. 6c,d) than for the extreme ones (Fig. 6a,b). The outflow area of the minor compound events in the South-West of Greenland coincides with the area of negative correlations in Fig. 4b. The outflow area of the top examples in the South-West of Iceland coincides with the positive correlations in Fig. 4b. Overall, the differences between the top and minor compound event trajectories suggest that the direction of outflow of rising air masses of warm conveyor belts influences downstream upper-level ridges. Accordingly, compounding extremes over Europe during omega-blocking are shown to be highly sensitive to far-upstream precursors emerging from synoptic weather systems.

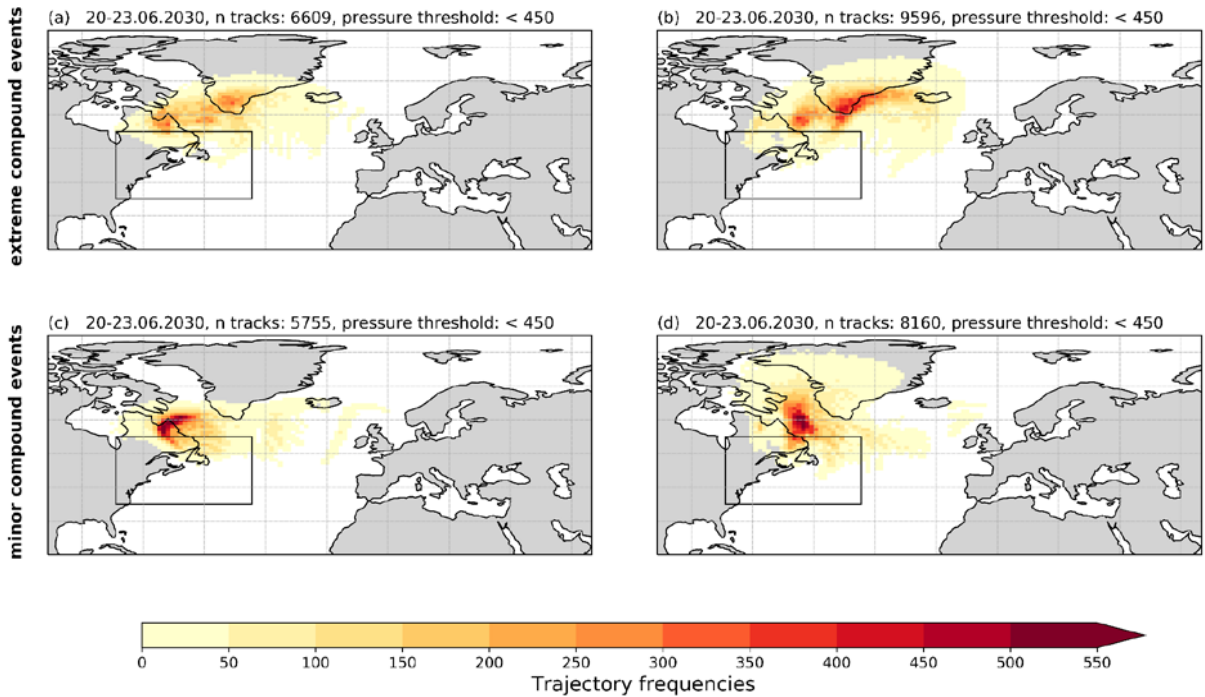

**Fig. 6 | Does the outstream of the warm conveyor belts differ between top and minor compound examples? a, b** Frequency of air parcel trajectories that have risen above 450 hPa per grid point for the period of June 20th-23rd 2030, for the two top examples. **c,d** same as a, b, but for two minor compound event examples from Fig. 5.

## Discussion

This study demonstrates the potential of ensemble boosting to systematically track dynamic questions along physically plausible event storylines within a fully coupled climate model. Leveraging this approach, we gain new insights into the large-scale atmospheric processes that govern the formation of omega-blocks with spatially compounding extremes, underscoring the critical influence of upstream cyclone dynamics and associated moist processes. We find that omega-blocks with compound extremes over Europe are highly sensitive to remote drivers upstream over the Atlantic. This applies both to the intensity and location of the omega block and above all to the intensity of the corresponding extremes (see Fig. S6-9). Days ahead of the extreme event, minor differences in the controlled experimental setup of ensemble boosting that are hardly visible on a weather map are identified, which later lead to immense differences in the resulting extremes over Europe. Correlation maps showing the correlation coefficients of heat extremes on peak date to the large-scale 500hPa geopotential height anomaly field demonstrate that up to five days before the event peak, "centers of action" with correlations >0.5 over the Atlantic exist. Air parcel tracking for selected simulations representing the top and minor compound event cases of the boosted ensemble reveals that in both groups, warm conveyor belt activity occurs in an upstream cyclone and that the areas involved in these moist processes correspond to the previously identified "centers of action". Warm conveyor belt activity differs little in intensity between the two groups, but there are significant differences in the outflow area and orientation of the warm conveyor belt. While minor compound event cases show a northerly outflow with a tendency to a more cyclonic outflow branch, top compound event members have a more anticyclonically-oriented warm conveyor belt outflow to the North-East, leading to an intensification of the downstream upper-level ridge. This suggests that the orientation of the WCB outflow branch is not only accompanied by different WCB characteristics[48] but also might affect the impact on the jet and downstream flow evolution. These differences in WCB-outflow indicate that compound extremes over Europe during omega-blocks are highly sensitive to far-upstream precursors. Observed links between WCB activity and North Atlantic sea surface temperatures suggest that ocean–atmosphere interactions can modulate the strength and orientation of upstream cyclones[49,50]. Given the high sensitivity of omega-blocks to WCB activity, such modulation could influence the subsequent development of downstream compound extremes under further global warming.

Our analysis is based on a model reference event rather than the September 2023 event itself. And the reference is simulated with one single model, as CESM2 offers a uniquely suitable framework for ensemble boosting and enables a clean experimental setup to investigate dynamical questions. The CESM2 model can reproduce omega-blocks with spatial patterns similar to the September 2023 event, and different reference events show comparable behaviour[43]. Using ensemble boosting, CESM2 can also generate omega-blocks with stronger spatially compounding extremes than observed. However, CESM2 tends to underestimate the frequency of spatially compounding extremes[43]. Previous studies have shown the capability of models to reproduce WCB activity for operational reforecasts[51] and for the predecessor model of CESM2, CESM[17]. Ref. 17 find that both dry and moist processes are reproduced remarkably well in a large ensemble of CESM compared to ERA-Interim. While climate models generally still struggle to reproduce blocking frequencies over Europe, there is a general tendency of improvement in the CMIP6 generation[52]. The study by ref. 27 analyzes eight CMIP6 models and argues that an underestimation of WCB outflow in climate models compared to ERA5 is, besides a misrepresentation of the mean flow, one of the key model biases that leads to an underestimation of blocking frequencies in the Euro-Atlantic sector.

Our findings are based on a single model-based case study, which may constrain the generality of the results. Future research could extend this dynamically conditioned approach to other fully coupled climate models or reforecast models to test the robustness of our findings on WCB outflow areas and their impact on upper-level ridge intensification downstream across different modeling contexts. Applying correlation maps and air parcel tracking allows us to identify the sensitivity to synoptic activity in upstream regions. In future research on extreme event storylines of boosted ensembles, a refined view on sensitive regions could emerge from diagnostics developed for the investigation of uncertainty growth in numerical weather prediction. E.g., ensemble sensitivity analysis[53] has been used to reveal the sensitivity of ridge amplification on forecast uncertainty in upstream WCB activity[54]. Ref. 55 suggests that moist processes may exert contrasting influences on different blocking regimes, with latent heat release in upstream cyclones enhancing block persistence, while in downstream cyclones it acts to weaken the block. Future work could address the question of whether similar mechanisms operate in omega-blocks, where the upstream and downstream lows may have opposing effects on block maintenance. Preliminary trajectory analyses for our case suggest limited ascent from the upstream cut-off low into the block core, implying a weak influence compared to the far-upstream cyclone (not shown).

Besides the limitations of our study to a single model and case, our findings demonstrate once more the extreme sensitivity of regional extreme events in the mid-latitudes to atmospheric dynamics. In the context of climate change, it is not only important to understand the mechanisms leading to extreme compound events but also why a moderate or weak event differs from an extreme one, as this understanding is key to making statements about how such events may change in a warming climate. A better understanding of the key processes governing omega-blocks over Europe and their representation in climate models is therefore essential for improving projections of spatially compounding extremes, which is especially critical because they hit a continent at several locations at the same time and thus can stress joint adaptation and risk-management efforts.

Our study additionally highlights that physically plausible storylines from ensemble boosting offer a powerful framework to: (1) advance our understanding of the processes driving e. g., compound extremes under omega-blocking, (2) track dynamic questions that require a clean experiment setup, (3) evaluate the capability of fully coupled climate models to reproduce these processes, and (4) contribute to improving the reliability of climate impact assessments for events like the European extremes in September 2023 through an improved understanding of the dynamic drivers.

## Materials and Methods

**Observations**

We use ERA5 reanalysis[56,57] of geopotential height at 500 hPa (z500; hourly), maximum 2m air temperature (daily), and total precipitation (daily) to describe the omega-blocking event with spatially compounding extremes that happened in September 2023 over Europe. The spatial resolution is 0.25°. Anomalies are calculated compared to the climatological mean of the period from 2005 to 2024 (31-day running mean for z500; 15-day running mean for tasmax).

**Ensemble Boosting**

To obtain a large database of omega-blocking situations, we employ the *Community Earth System Model version 2 (CESM2*[42]*)* in combination with an event-based re-initialization method called ensemble boosting[39-41]. For this purpose, we first select a model-based reference event (referred to as unperturbed) from a 30-member large ensemble of the CESM2 model (CESM2-LE). We then slightly perturb the initial conditions 10 to 18 days before the peak date of the reference event on June 28th (starting dates: 11-18th June 2030) by introducing round-off errors ($10^{-13}$) in the specific humidity field on each grid point. We produce 100 simulations for these eight different lead times. This way, we obtain an event-based "mini-ensemble" of 800 coherent physical event trajectories for potential omega-blocking events with compound extremes. This experimental design creates a controlled set of event-based simulations, all initialized from the same large-scale atmospheric state and boundary conditions (e.g., sea surface temperature, soil moisture), thereby enabling a robust process-oriented analysis of atmospheric evolution along physically plausible event storylines.

**Selection of reference event**

The unperturbed reference event for ensemble boosting is selected using a ranking-based metric as described in ref. 43. This metric is designed for the identification of omega-patterns, considering the compounding nature of the flood-heat-flood events. The ranking-based metric uses three preset boxes over the three impact regions over Europe during the event in September 2023: Iberian Peninsula, Northern France, and Greece (see Fig. 1). The box definitions are based on a search for events that are comparable to September 2023 both in terms of atmospheric flow and heat/precipitation extremes, while allowing some flexibility in the area of occurrence (see ref. 43 for details). We use this joint ranking metric to select time steps in the CESM2-LE with the lowest rank for all six variables: the atmospheric omega-pattern (low-pressure over Iberia, high-pressure over Northern France, and low-pressure over Greece in z500) and the compound extremes (precipitation extremes over Iberia and Greece, and heat extremes over Northern France). From five potential reference events, we choose the one with the strongest grid-cell level precipitation for a detailed analysis of atmospheric conditions in this study (corresponding to case 2 in ref. 43).

**Selection of top and minor compound event groups**

To compare ensemble members with strong versus weak compound extremes, we again apply a rank-based metric as for the selection of the reference event[43], only this time we exclude pressure patterns from the joint ranking to be able to independently examine how atmospheric circulation differs between the two groups. The criteria for the ranking are: daily precipitation over Iberia and Greece, and daily maximum temperature (tasmax) over Northern France. The joint ranking metric considers the maximum values of absolute daily precipitation and maximum daily temperature (tasmax) within a five-day moving window after smoothing with three-day running means (5-day max. of Tx3day/Rx3day). The analysis focuses on a 15-day window centered on the peak date of the reference event. This prevents the selection of unrelated compound extremes that occur late in the 50-day time series and are unrelated to the targeted omega-block. We compare the strongest and weakest 10% of the eight starting days. While the strongest 10% of compound events all show an omega-pattern with compound extremes over the target regions, the 10% weakest cases from the ranking build a more heterogeneous group: the majority has an omega-pattern with weak compound events over the target regions, and others having no omega-pattern and instead other blocking types (e.g., dipole structure), or an (displaced) omega-pattern but without three compound events. There are a few exceptional cases that show a spatially shifted omega-pattern with compound extremes, which are not captured by the target regions.

We use an anomaly-based blocking index from the Python package ConTrack[58] to check if the anticylones of the members from the top event group indeed count as a block and if their position lies over the Northern France box. The minor compound event group might include members without a block over Northern France. The blocking index tracks contours of anomalous pressure in the z500 field (above the 90th percentile) and classifies an anticyclone as 'blocked' if it persists for at least five days (persistence criteria) with a spatial overlap of the contours of >= 50 %, which ensures slow movement of the anticyclone (velocity criteria). The z500 anomalies are calculated relative to the 31-day running mean climatology of the 30-member CESM2-LE in the years 2005 to 2024 for the summer half-year (JJASON) in the mid-latitudes (50-80°N).

**Lagrangian air parcel tracking for selected boosted members**

We trace air parcels of selected members from the boosted ensemble to identify trajectories that fulfill the criteria for warm conveyor belts. For the air parcel tracking, zonal-(U), meridional-(V), and vertical-(Ω) wind components are employed in 6-hourly resolution. This requires a re-calculation of the selected boosted members to save these 3D wind variables in high temporal resolution (h3-files in CESM2 file formatting). We select four members for re-boosting (markers in Fig. 2): 1. the member with the strongest compound extremes based on the joint-rank (lowest rank of all 800 members = TOP1) and a corresponding counter-part from the minor compound event group, which is the member with the weakest compound extremes (highest joint rank) of the same lead time (June 13th; -16 days before peak date of reference event), and 2. the member with the second strongest compound extremes from a different lead time, which happens to be the top third with a lead time of -17 days (June 12th) and its corresponding weakest member from the same lead time from the minor compound event group.

We use the software ETH-Lagranto[47] (CESM2 version) and calculate 48h forward trajectories from a preset starting box (-80 to -40°W, 35 to 55°N) that covers the area of the upstream cyclone with equidistant starting points every 100 km. Trajectories can start between 1000 and 700 hPa. For CESM2, there are 32 vertical levels available (992.6hPa to 3.6hPa). From all tracked trajectories, only those are selected that fulfill the criterion for warm conveyor belts with an uplift of at least 400hPa in 48h.

**Acknowledgements**

**Funding**

M.M. acknowledges funding from the LMUexcellent funding line at LMU Munich. E.F., E.B., and J.Z. acknowledge funding from the European Union's Horizon 2020 research and innovation programme under grant agreement no. 101003469 (XAIDA). E.B. received funding from the DFG via the Emmy Noether Programme (Grant ID 524780515). L.S.G. received funding from the European Union's Horizon Europe Framework Programme under the Marie Skłodowska-Curie grant agreement No 101064940. Y.G. acknowledges support from the China Scholarship Council Programme (grant ID: 202306010362) and the National Natural Science Foundation of China (Nos. 42175065 and 41975059).

**Author contributions**

M.M., C.G., and E.F. designed the study. U.B. performed the climate model experiments. M.M. performed the data analysis, produced the figures and led the writing. Y.G. performed the calculation of the ranking based metric for the reference event. C.G., L.S.G., E.B., J.Z. and E.F. contributed to the interpretation of the results and the writing of the paper.

**Competing interests**

The authors declare that that one author is part of the editorial team at Science Advances.

**Data and materials availability**

The ERA5 reanalysis data used in this study are available from the European Centre for Medium-Range Weather Forecasts (ECMWF), Copernicus Climate Change Service (C3S) at Climate Data Store (CDS; https://cds.climate.copernicus.eu/). The processed CESM2 simulations of the boosted ensemble are available via the Zenodo repository[59] at https://zenodo.org/records/15644833. The code used to generate the figures in this paper and the Supplementary Material is available from github: https://github.com/mmittermeier/OmegaBoosting.git.

**Supplementary Material**

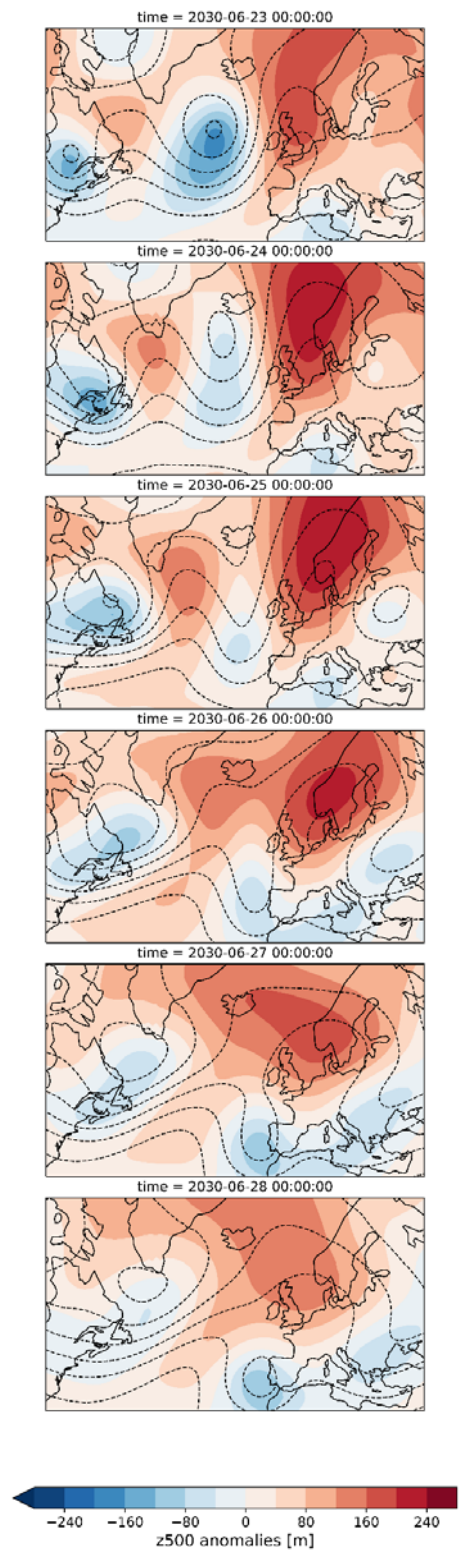

**Fig. S1 | Composite maps of the 500hPa geopotential height field of the minor compound event group.** Same as Fig. 3 but averaged over the minor compound event group. Colors show z500 anomalies. Absolute z500 values are indicated by dashed contours.

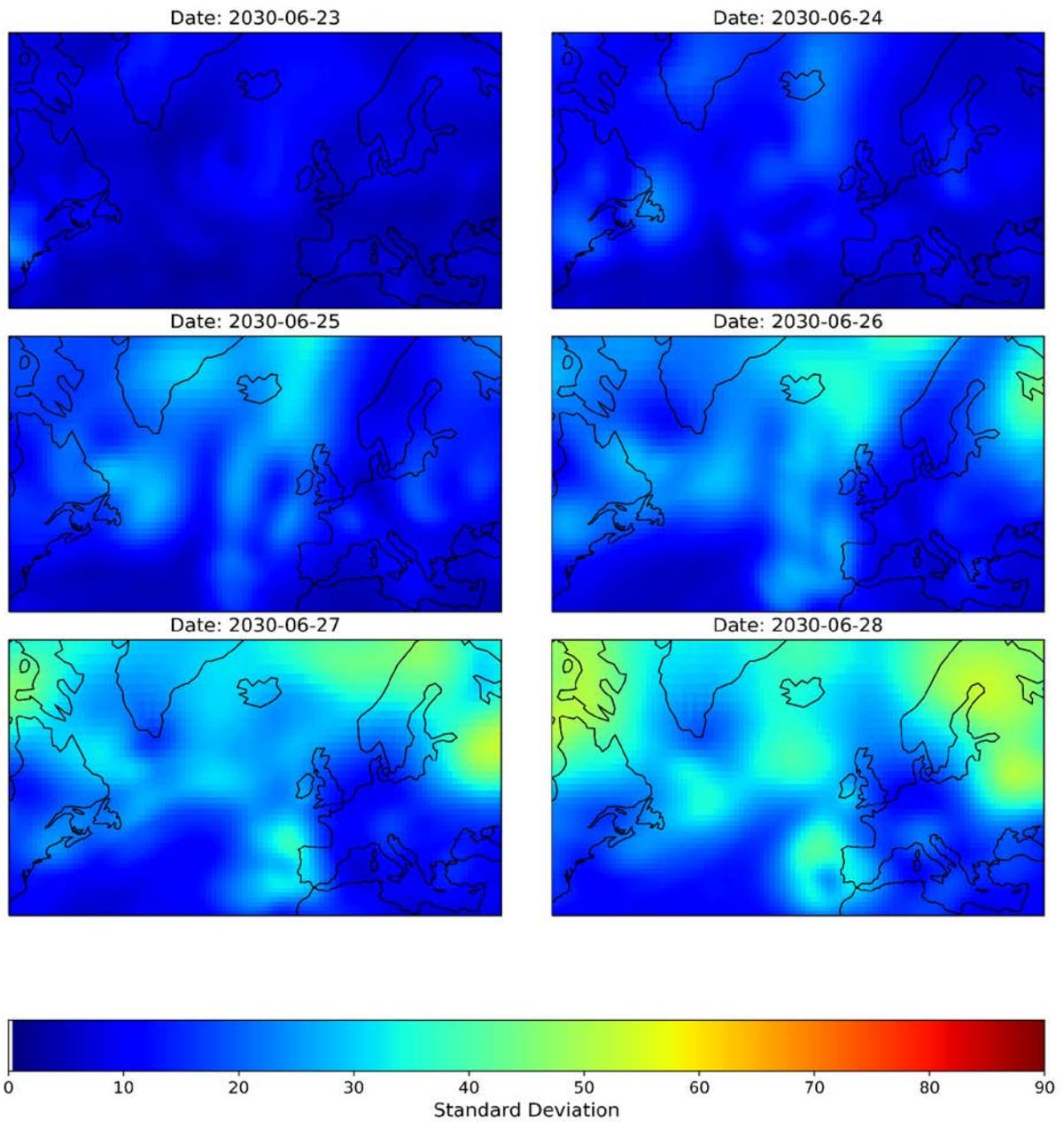

**Fig. S2 | Standard deviation of z500 anomalies of the top compound event group.** Standard deviation in meters [m] of the 500hPa geopotential height anomalies during the onset and peak of the omega-pattern.

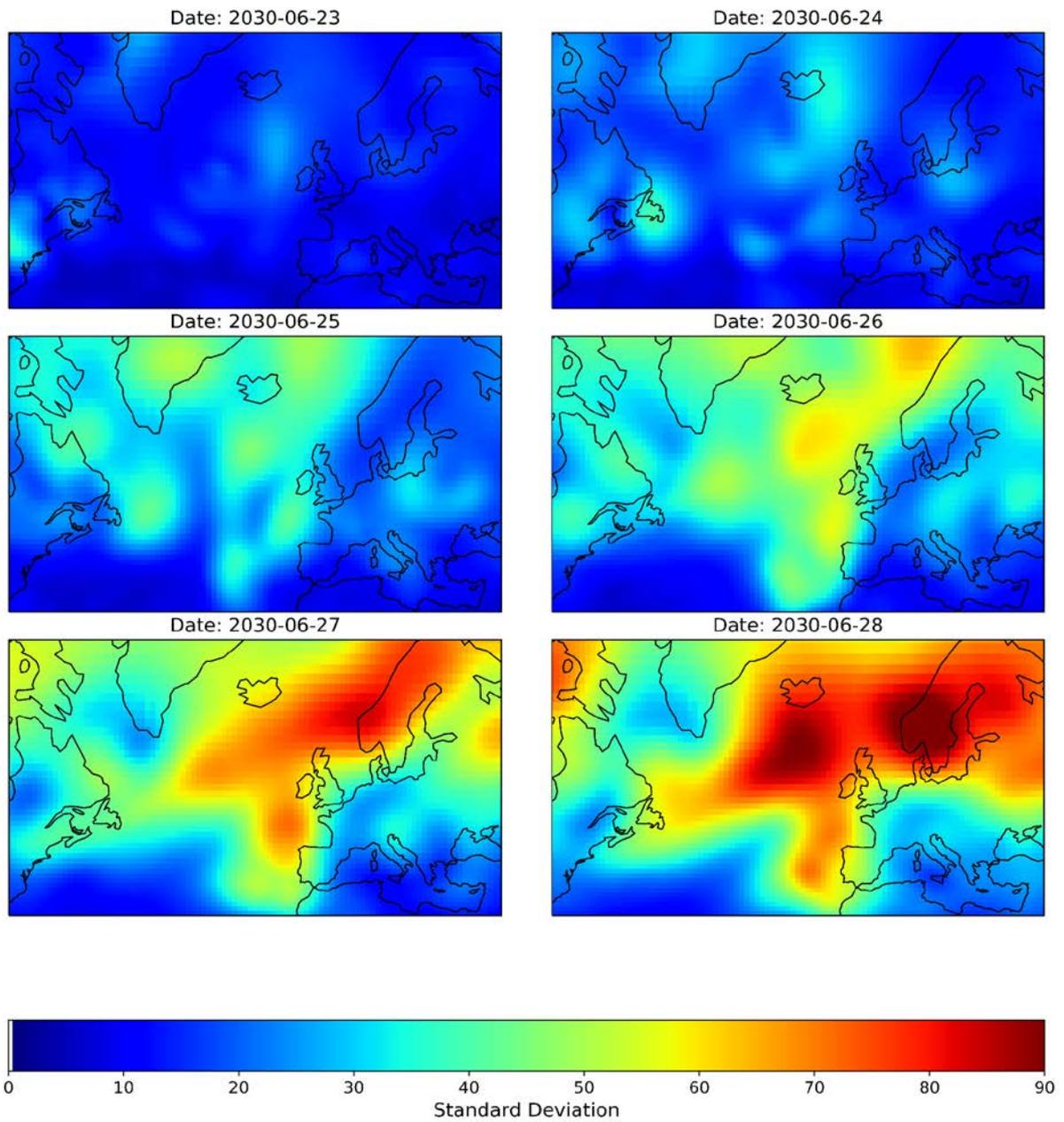

**Fig. S3 | Standard deviation of z500 anomalies of the minor compound event group.** Same as S2 but for the minor compound event group.

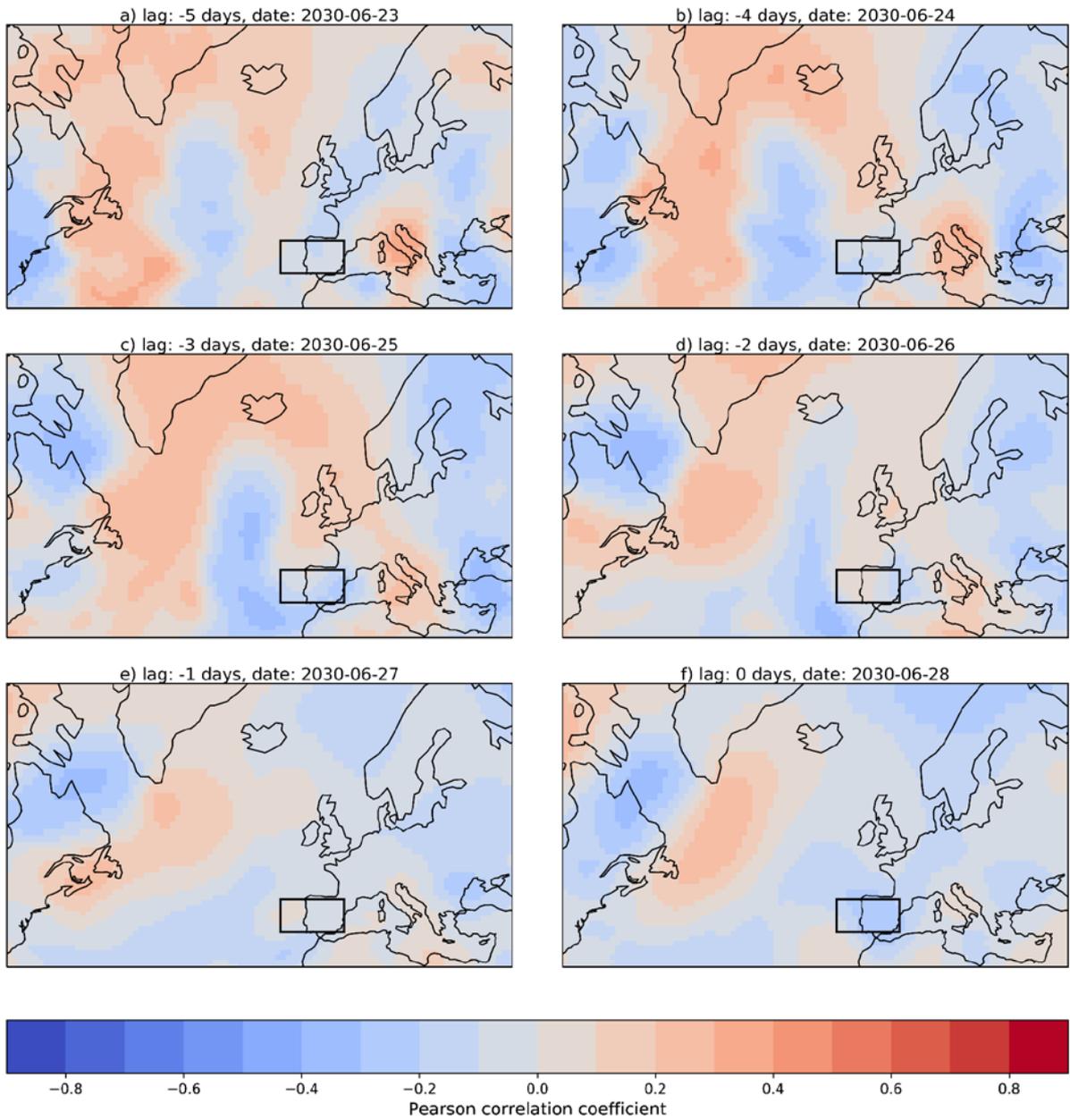

**Fig. S4 | Correlation maps for precipitation over the Iberian Peninsula.** Same as Fig. 4, but for precipitation over the Iberian Peninsula: Pearson correlation coefficients between daily precipitation sum averaged over the 40% hottest pixels in the Iberian Peninsula (outlined box) on the event peak day of the reference event (2030-06-28), and 500 hPa geopotential height (z500) anomalies at each grid point across the domain, for various time lags.

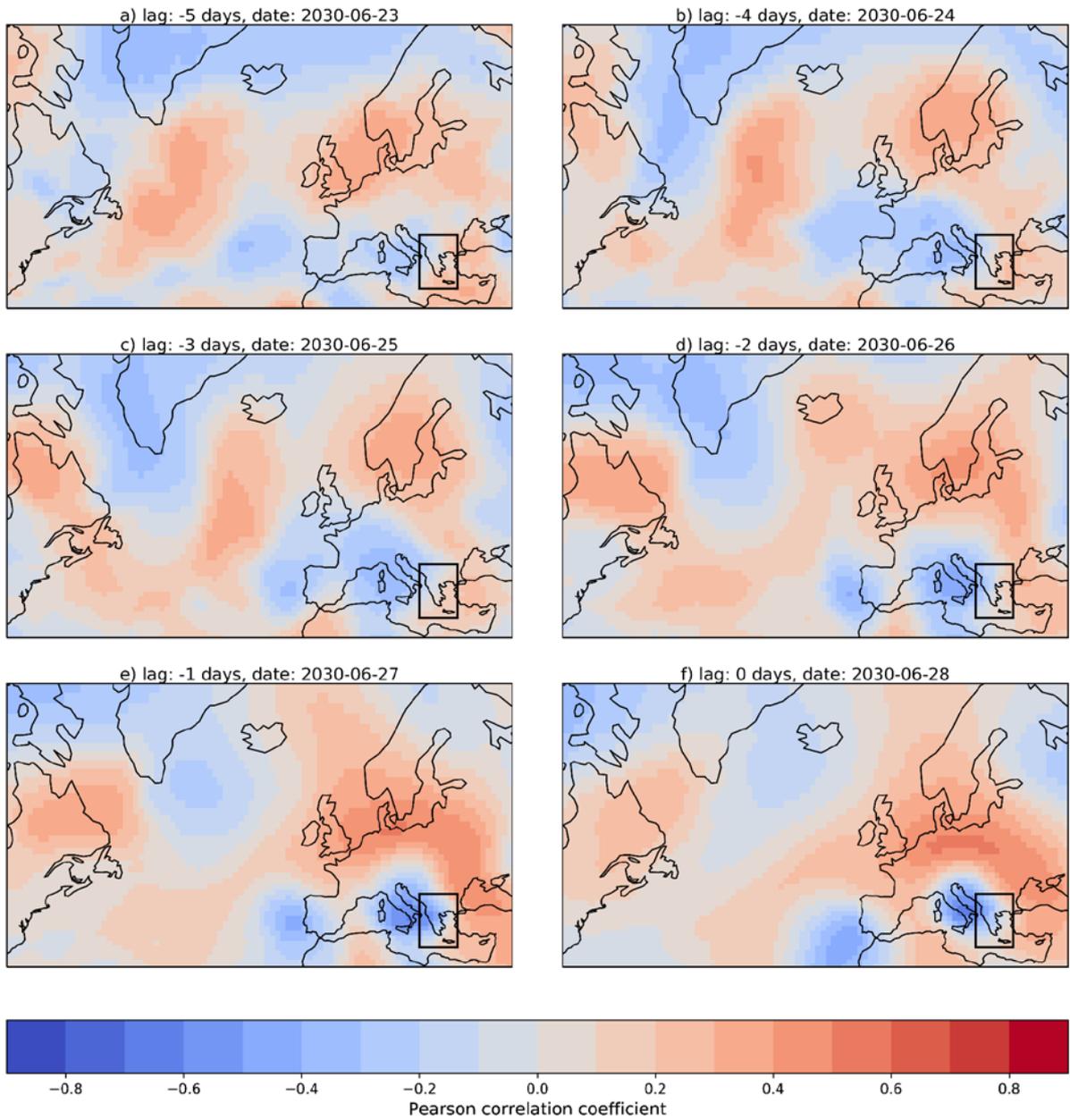

**Fig. S5 | Correlation maps for precipitation over Greece.** Same as Fig. S1, but for daily precipitation sum averaged over the 40% wettest pixels in Greece (outlined box).

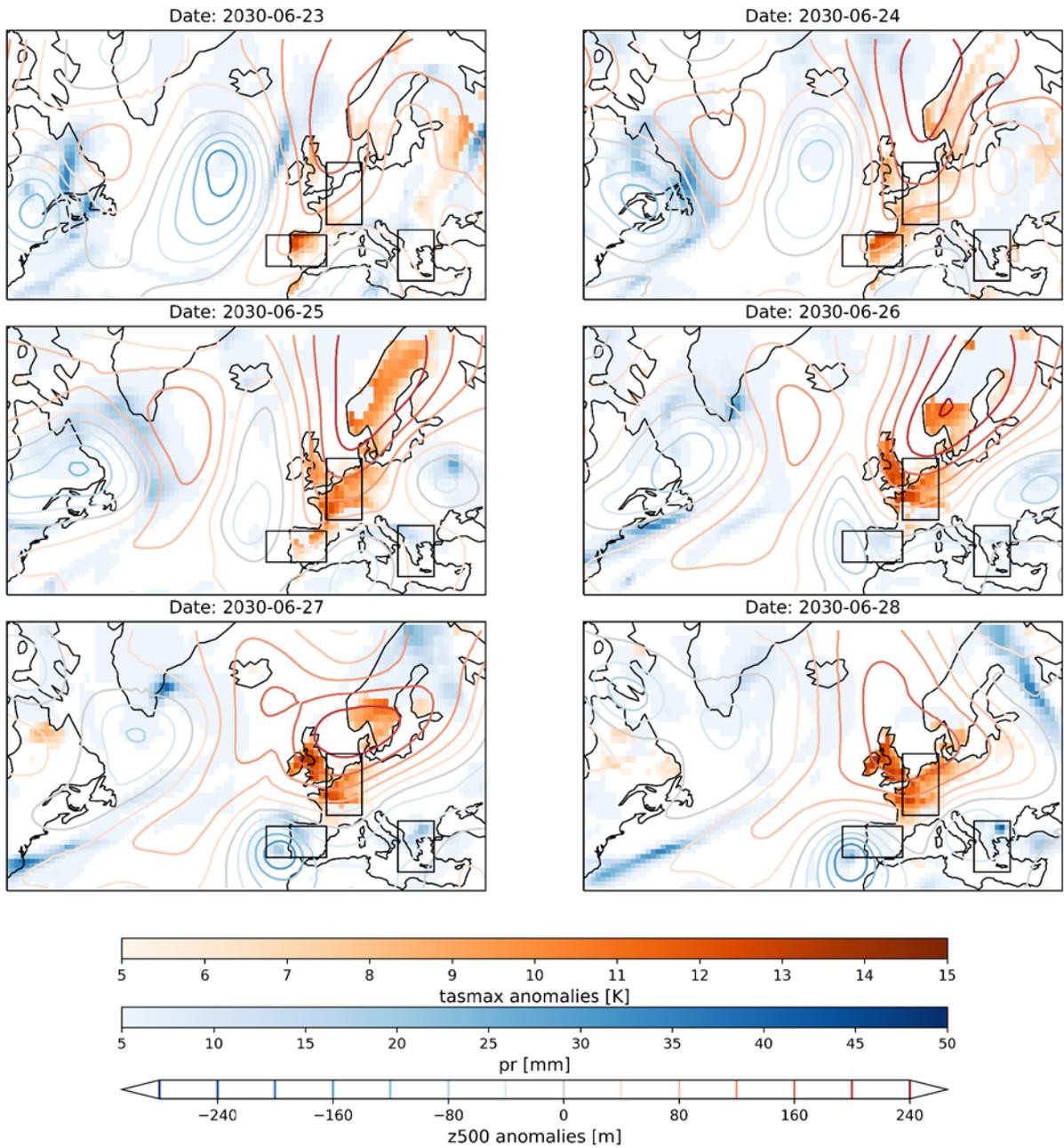

**Fig. S6 |Omega-Pattern and spatially compounding extremes for the top ranking member (TOP1)** (a) averaged 500 hPa geopotential height anomalies (contours), averaged temperature anomalies (red; only plotted for anomalies > 5K; threshold chosen for readability purposes), and precipitation sums (only plotted for pixels with pr > 5 mm) from 22nd to 29th June 2030. Anomalies are calculated compared to 2005-2024 (31-day running mean). Temperature anomalies are clipped with the land-sea mask. This member corresponds to the one shown in Fig. 4a and Fig. 5a.

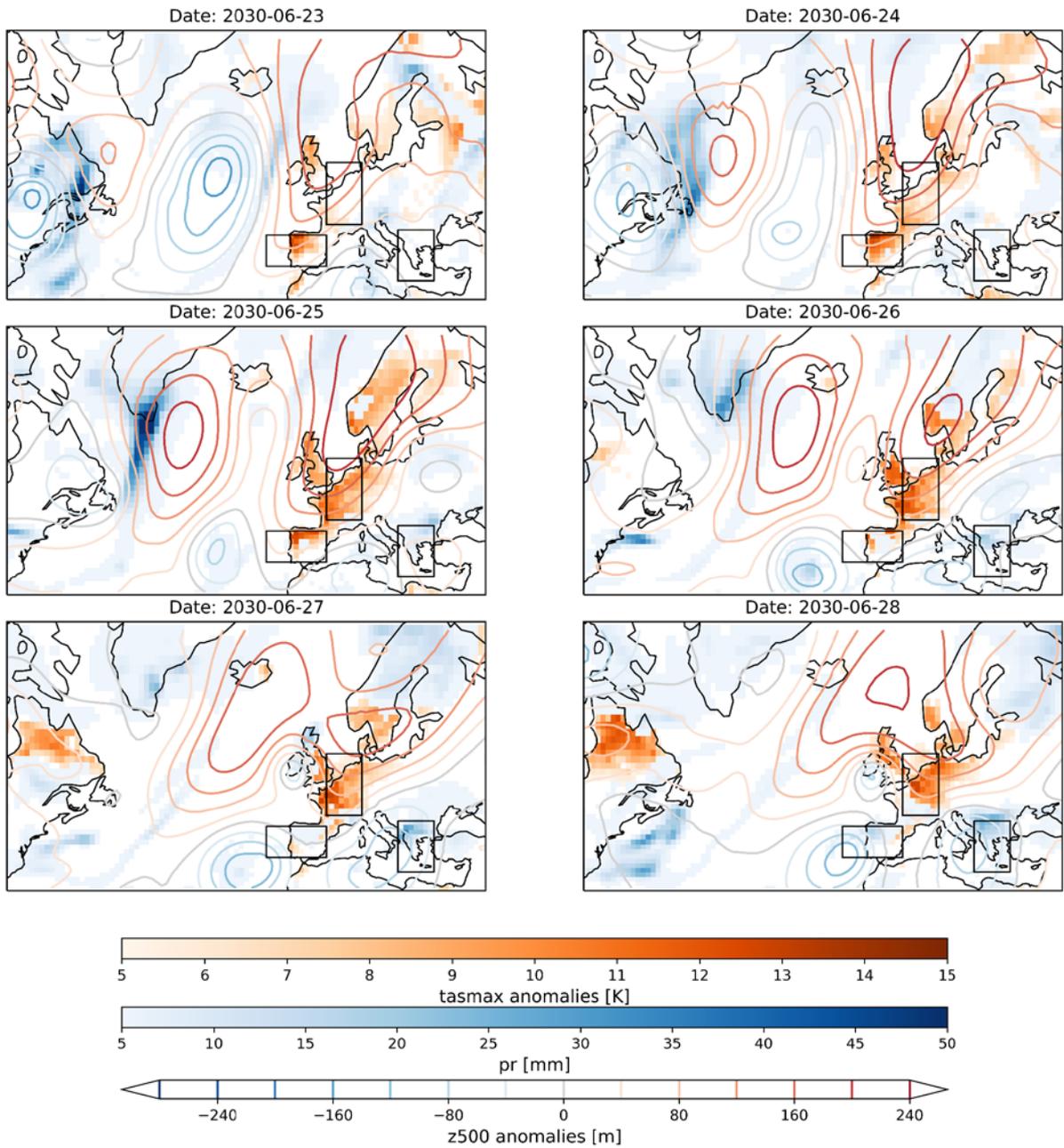

**Fig. S7 | Omega-Pattern and spatially compounding extremes for the third highest ranking member overall (TOP3), representing the second highest at a different lead time than TOP1** (a) averaged 500 hPa geopotential height anomalies (contours), averaged temperature anomalies (red; only plotted for anomalies > 5K; threshold chosen for readability purposes), and precipitation sums (only plotted for pixels with pr > 5 mm) from 22nd to 29th June 2030. Anomalies are calculated compared to 2005-2024 (31-day running mean). Temperature anomalies are clipped with the land-sea mask. This member corresponds to the one shown in Fig. 4c and Fig. 5c.

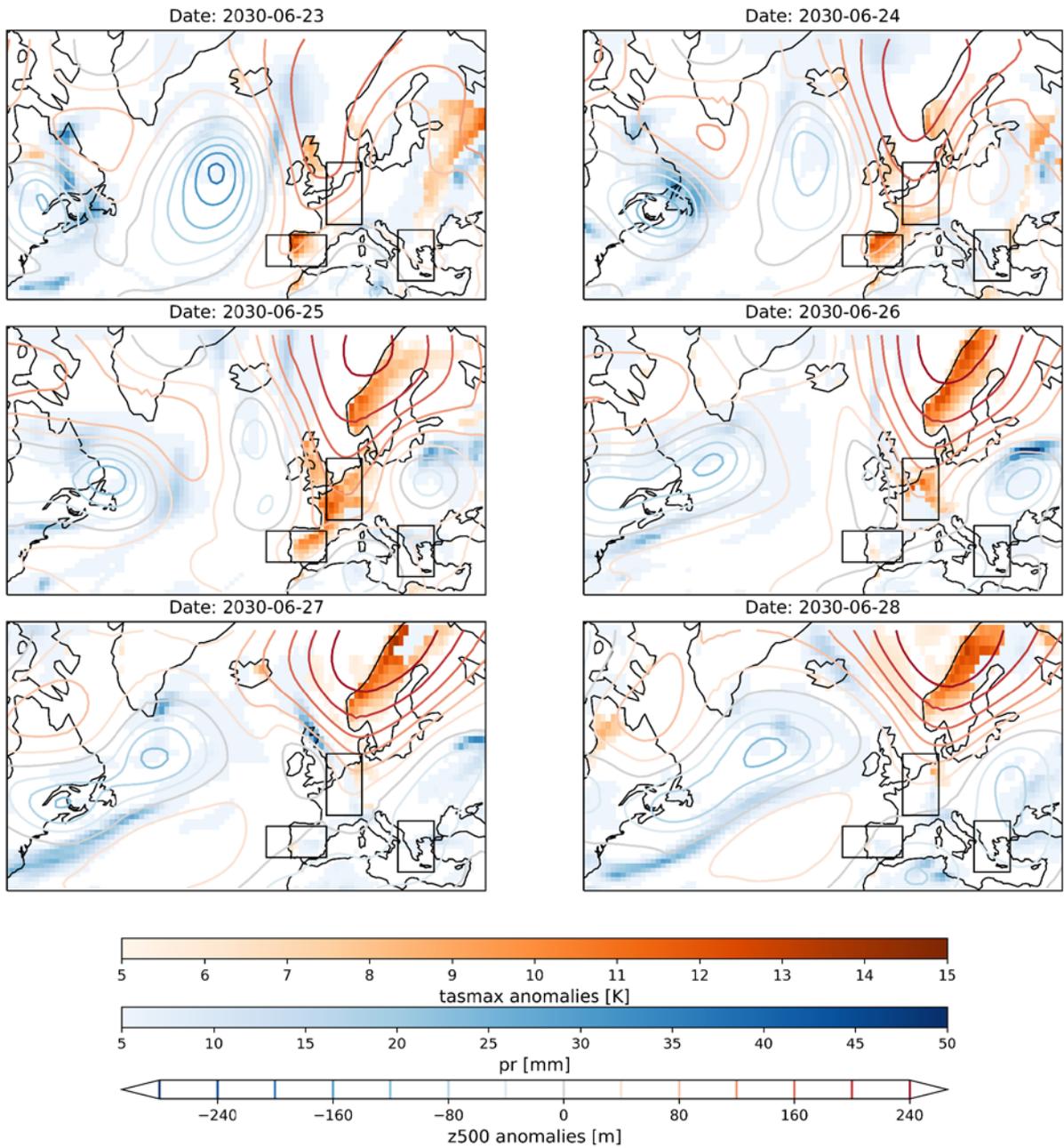

**Fig. S8 |Z500 anomaly field, temperature, and precipitation over Europe for the member with the weakest compound events from the same lead time than TOP1** (a) averaged 500 hPa geopotential height anomalies (contours), averaged temperature anomalies (red; only plotted for anomalies > 5K; threshold chosen for readability purposes), and precipitation sums (only plotted for pixels with pr > 5 mm) from 22nd to 29th June 2030. Anomalies are calculated compared to 2005-2024 (31-day running mean). Temperature anomalies are clipped with the land-sea mask. This member corresponds to the one shown in Fig. 4b and Fig. 5b.

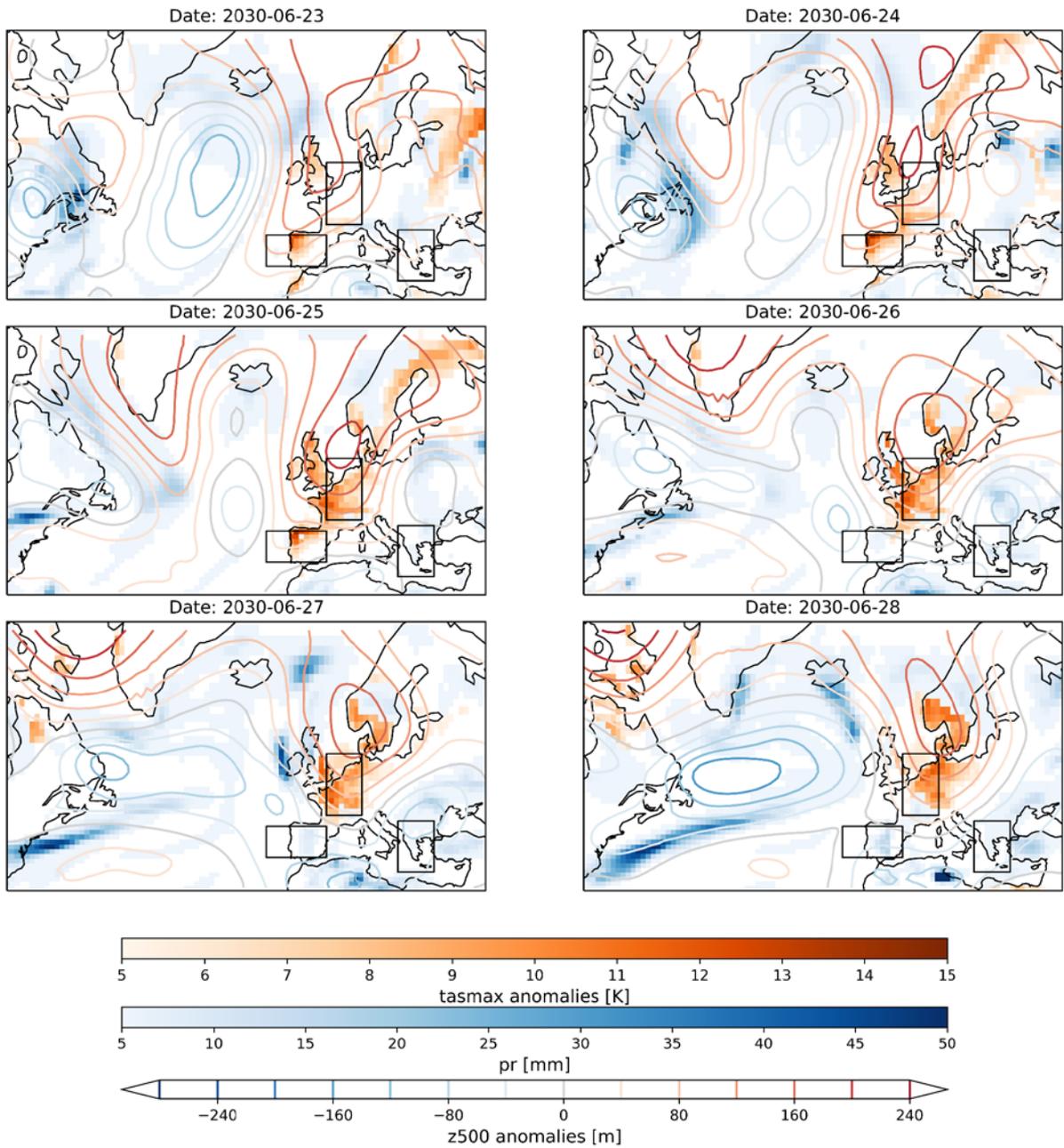

**Fig. S9 |Z500 anomaly field, temperature, and precipitation over Europe for the member with the weakest compound events from the same lead time than TOP3** (a) averaged 500 hPa geopotential height anomalies (contours), averaged temperature anomalies (red; only plotted for anomalies > 5K; threshold chosen for readability purposes), and precipitation sums (only plotted for pixels with pr > 5 mm) from 22nd to 29th June 2030. Anomalies are calculated compared to 2005-2024 (31-day running mean). Temperature anomalies are clipped with the land-sea mask. This member corresponds to the one shown in Fig. 4d and Fig. 5d.